\documentclass[aps,prl,reprint,showpacs,showkeys]{revtex4-1}
\usepackage{microtype}
\usepackage{amsthm}
\usepackage{amsmath}
\usepackage{amsfonts}
\usepackage[mathscr]{eucal}
\usepackage{amssymb,epsfig,setspace}
\usepackage{graphicx}
\usepackage{natbib}
\begin{document}
\title[How to calculte spectra?]{How to calculte spectra of Rabi and
  related models?}%
\author{Andrzej J.~Maciejewski} \email{maciejka@astro.ia.uz.zgora.pl}
\affiliation{J.~Kepler Institute of Astronomy, University of Zielona
  G\'ora, Licealna 9, PL-65--417 Zielona G\'ora, Poland.}%
\author{Maria Przybylska}%
\email{M.Przybylska@proton.if.uz.zgora.pl} \affiliation{ Institute of
  Physics, University of Zielona G\'ora, Licealna 9, 65--417 Zielona
  G\'ora, Poland }%
\author{Tomasz Stachowiak} \email{stachowiak@cft.edu.pl}
\affiliation{%
  Center for Theoretical Physics PAS, Al. Lotnikow 32/46, 02-668
  Warsaw, Poland }%

\date{\today}%
\begin{abstract}
  We show how to use properly the Bargmann space of entire functions
  in the analysis of the Rabi model. We are able to correct a serious
  error in recent papers on the topic and develop a corrected method
  of finding the spectrum applicable also to other, more general
  systems.
\end{abstract}
                      
\maketitle


Several papers devoted to determination of the spectrum of the Rabi
model have appeared recently, see e.g.
\cite{Braak:11::,Braak:12::a,Moroz:12::}, and references therein for a
description of various approaches to this problem. Let us recall that
the Rabi model describes interaction of a two-level atom with a single
harmonic mode of the electromagnetic field.  In the Bargmann-Fock
representation this model is described by the following system of two
differential equations
\begin{equation}
  \begin{split}
    (z+\lambda)\dfrac{\mathrm{d}\psi_1}{\mathrm{d}z}=&(E-\lambda
    z)\psi_1-\mu\psi_2,\\
    (z-\lambda)\dfrac{\mathrm{d}\psi_2}{\mathrm{d}z}=&(E+\lambda
    z)\psi_2-\mu\psi_1,
  \end{split}
  \label{eq:syst}
\end{equation}
where $E$ is the energy, $\lambda$ is the atom-field coupling
constant, and $2\mu$ is the atomic level separation.  In this
representation, two component wave function $\psi=(\psi_1, \psi_2)$ is
an element of Hilbert space $\mathscr{H}^2=\mathscr{H}\times
\mathscr{H}$, where $\mathscr{H}$ is the Bergmann-Fock Hilbert space
of entire functions of one variable $z\in\mathbb{C}$. The scalar
product in $\mathscr{H}$ is given by
\[
\langle
f,g\rangle=\dfrac{1}{\pi}\int_\mathbb{C}\overline{f(z)}g(z)e^{-|z|^2}\mathrm{d}
(\Re(z))\mathrm{d}(\Im(z)).
\]
An entire function $f(z)$ belongs to $\mathscr{H}$ if it has the
proper growth at infinity, for details, see \cite{Bargmann:61::}.
Thus, energy $E$ belongs to the spectrum of the problem, if and only
if for this value of $E$ equations~\eqref{eq:syst} have entire
solution $\psi= (\psi_1, \psi_2)$ with the proper behaviour at
infinity.  However, for the considered system all entire solutions
belong to $\mathscr{H}^2$, see \cite{Schweber:67::} for an
explanation.

Equations~\eqref{eq:syst} have two singular regular points at
$z=\lambda$ and $z=-\lambda$, while infinity is an irregular singular
point.  The system has a $\mathbb{Z}_2$ symmetry. It is invariant with
respect to the involution $\tau:\mathscr{H}^2 \rightarrow
\mathscr{H}^2$ given by $ \tau(\psi_1, \psi_2)(z)= (\psi_2(-z),
\psi_1(-z))$. In other words, if $ (\psi_1(z), \psi_2(z))$ is a
solution of this system, then also $ (\psi_2(-z), \psi_1(-z))$ is its
solution.  We say that a solution $\psi=(\psi_1, \psi_2)$
of~\eqref{eq:syst} has parity $\sigma\in\{-1, +1\}$, if
$\tau(\psi)=\sigma \psi$.

In \cite{Braak:11::} it was pointed out that the the above mentioned
symmetry plays the crucial role in determination of its spectrum.
Later, in \cite{Moroz:12::}, it was shown that one can determine the
spectrum of the Rabi model without any references to its
$\mathbb{Z}_2$ symmetry. However, the method used in~\cite{Moroz:12::}
has limited application as it is based on a result valid only for
third order recurrence relations, see \citep{Gautschi:67::}.
 
Braak's approach is very interesting because it gives a chance for
generalisation beyond third order recurrence relations. However in its
original formulation it contains a flaw which invalidates its results
concerning the spectrum of the standard Rabi model. The aim of our
paper is to correct results of~\cite{Braak:11::} and to present our
direct method which can be used to study of systems more complicated
than the Rabi model.
 
We applied our method to study a generalised Rabi model with broken
$\mathbb{Z}_2$ symmetry introduced in \cite{Braak:11::}.  In the
Bargmann representation it is described by the following equations
\begin{equation}
  \begin{split}
    &(z+\lambda)\dfrac{\mathrm{d}\psi_1}{\mathrm{d}z}= (E-\epsilon
    -\lambda
    z)\psi_1-\mu\psi_2,\\
    &(z-\lambda)\dfrac{\mathrm{d}\psi_2}{\mathrm{d}z}=
    (E+\epsilon+\lambda z)\psi_2-\mu\psi_1.
  \end{split}
  \label{eq:syste}
\end{equation}
For $\epsilon=0$ it coincides with the Rabi model~\eqref{eq:syst}. We
denote by $p:=(E,\lambda,\mu,\epsilon)$ parameters of this
model. Instead of $E$ we will use also the parameter $x:=E+\lambda^2$.
Let us assume that $x$ is not a non-negative integer.  Then the
spectrum of the problem coincides with zeros of one function $w(p)$
defined in the following way. Let
\begin{equation}
  \label{eq:w}
  w(p;y):= H_1(y)H_2'(y) - H_1'(y)H_2(y),
\end{equation}
where
\begin{equation}
  \label{eq:hc}
  H_1(y):=\operatorname{HeunC}(a_0
  ; y), 
\end{equation}
and
\begin{equation}
  \label{eq:hc}
  H_2(y):=\operatorname{HeunC}(a_1;1- y), 
\end{equation}
are the confluent Heun functions, see~\cite{Ronveaux:95::}, with
parameters $a_0:=(\alpha,\beta,\gamma,\delta,\eta) $, and
$a_1:=(-\alpha,\gamma,\beta,-\delta,\delta+\eta) $; in terms of
$(x,\lambda,\mu,\epsilon)$ these parameters are given by
$\alpha=4\lambda^2$, $\beta=-x+\epsilon$, and
\begin{equation}
  \begin{split}
    &\gamma=-1-x-\epsilon,
    \qquad\delta= 2(1-2\epsilon)\lambda^2,\\
    &2\eta= 1 - 2 \mu^2 + (1+x)(x - 4 \lambda^2 )+ \epsilon (1 + 4
    \lambda^2)-\epsilon^2.
  \end{split}
  \label{eq:par_heun_def}
\end{equation} 
Then, $W(p):=w(p;1/2)$.

Here we just recall the basic steps of the procedure which was
formulated for the Rabi model~\eqref{eq:syst}. It is fully described
in~\cite{Braak:11:sup}.

\begin{figure}[ht]
  \begin{center}
    \includegraphics[scale=0.57]{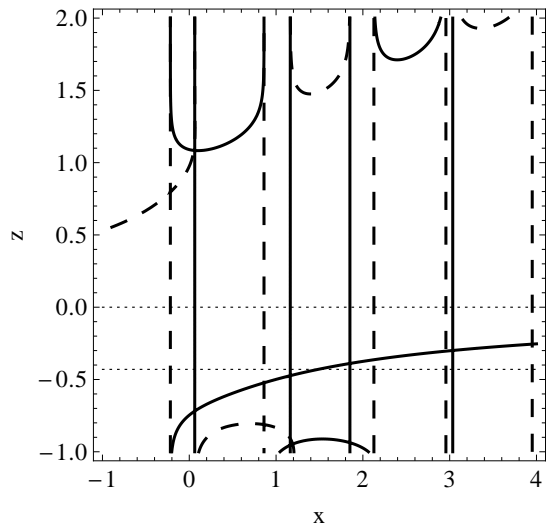}
  \end{center}
  \caption{\label{fig:1}The zero level contours of $G_+(z,p)$
    (continuous lines), and $G_-(z,p)$ (dashed lines) on the $(x,z)$
    plane, for $\lambda=0.7$ and $\mu=0.4$. }
\end{figure}

Let us introduce new variables $y:=z+\lambda$,
$\varphi_i(y)=\mathrm{e}^{-\lambda y}\psi_i(y-\lambda)$, for i=1,2.
It is obvious that functions $\psi_i(z)$ are entire if and only if
functions $\varphi_i(y)$ are entire.  In the new variables
system~\eqref{eq:syst} reads
\begin{equation}
  \begin{split}
    y\dfrac{\mathrm{d}\varphi_1}{\mathrm{d}y}=&x\varphi_1-\mu\varphi_2,\\
    (y-2\lambda)\dfrac{\mathrm{d}\varphi_2}{\mathrm{d}y}=&(x
    -4\lambda^2+2\lambda y)\varphi_2-\mu\varphi_1,
  \end{split}
  \label{eq:sysy}
\end{equation}
Let us assume that for a given $p=(x,\lambda,\mu)$,
system~\eqref{eq:sysy} has an entire solution
$\varphi(y)=(\varphi_1(y), \varphi_2(y))$, where
\begin{equation}
  \label{eq:ser2}
  \varphi_1(y)=\sum_{n=0} ^\infty\widetilde{\Phi}_n(p) y^n, 
  \quad \varphi_2(y)=\sum_{n=0} ^\infty \Phi_n(p) y^n.
\end{equation}
To proceed further we assume also that $x\not\in\mathbb{Z}$, since the
existence of solutions for integer $x$ has long been established, see
\cite{Kus:86::}. Then we insert the expression for $\varphi_2(y)$ into
the first equation~\eqref{eq:sysy}, and integrate it. We obtain
\begin{equation}
  \label{eq:PHi1}
  \widetilde{\Phi}_n(p)=\frac{\mu}{x-n} \Phi_n(p),\quad \text{for}\quad n\in\mathbb{N}. 
\end{equation}
Function $\Phi_n(y)$ can be determined recursively. In this way we
obtain a solution $\varphi(y)$, which is locally holomorphic at $y=0$.
Thus, we have a solution $\psi(z)$ which is locally holomorphic at
$z=-\lambda$. However, we are looking for an entire solution. In order
to find conditions which guarantee that $\psi(z)$ is entire we can
assume, without loss of generality that it has a given parity
$\sigma$. Thus putting $\psi_\sigma (z)= (\psi_{1,\sigma}(z),
\psi_{2,\sigma}(z))$ we have
$\psi_{1,\sigma}(z)=\sigma\psi_{2,\sigma}(-z)$ for all
$z\in\mathbb{C}$.  Hence, function
\begin{equation}
  \label{eq:G+}
  G_{\sigma}(z,p):=\psi_{2,\sigma}(-z)-\sigma\psi_{1,\sigma}(z), 
\end{equation}
must vanish identically for all $z\in\mathbb{C}$. In
\cite{Braak:11:sup}, Braak claims that
\begin{enumerate}
\item if $G_\sigma(z,p)=0$ for a fixed but \emph{arbitrary} $z$
  belonging to the intersection of domains of definiteness of
  $\psi_{1,\sigma}(z)$ and $\psi_{2,\sigma}(-z)$, then it  vanishes
  identically. That is, if for a given $p$ there exists $z_\star \in
  D(-\lambda, 2|\lambda|) \cap D(\lambda, 2|\lambda|)$ such that
  $G_\sigma(z_\star,p)=0$, then $G_\sigma(z,p)=0$ for an arbitrary
  $z\in \mathbb{C}$. Here $D(z_0, r) \subset\mathbb{C}$ is an open
  disc of radius $r$ and the centre at $z_0$.
\item $x$ belongs to the spectrum of the problem if and only if there
  exists $z_\star \in D(-\lambda, 2|\lambda|) \cap D(\lambda,
  2|\lambda|)$ such that $G_\sigma(z_\star,p)=0$, for a certain
  $p=(x,\lambda, \mu)$.
\end{enumerate}
Both of these statements are incorrect. There exist entire functions,
e.g., $\sin(z)$, which have infinite number of zeros but they are not
zero functions. So, the first claim needs at least some
justification. We show that in the considered context it is not valid.
To see this it is enough to make a contour graph of function
$G_{\sigma}(z,p)$ on the plane $(x,z)$, see Fig.~\ref{fig:1}.
One can notice that there exist $x$ such that $G_{\sigma}(z,p)=0$
identically in $z$, as well as that there exist isolated zeros of
$G_{\sigma}(z,p)$.  An example is presented in Fig.~\ref{fig:1}.  This
also shows that the second claim of Braak is not correct.

\begin{figure}[ht]
  \begin{center}
    \includegraphics[width=1\columnwidth,clip]{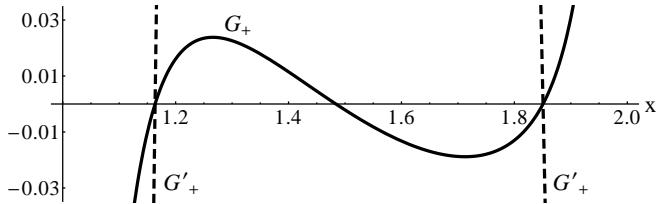}
  \end{center}
  \caption{\label{fig:2} Graphs of $G_+(z_\star,p)$ (continuous line),
    and $G'_+(z_\star,p)$ (dotted lines) for $z_\star=-0.43$ as
    functions of $x$, for $\lambda=0.7$ and $\mu=0.4$. The middle zero
    is simple (hence $G_+\not\equiv0$) and the correspoding $x$ is not
    in the spectrum.}
\end{figure}

We notice that the function $g(z):=G_\sigma(z,p)$, as a function of
$z$, satisfies a second order linear differential equation.  In fact,
it can be checked directly that
\begin{multline}
  \label{eq:g}
  (z^2-\lambda^2) g'' + \left[ z(1-2x -3\lambda^2)-\lambda\right] g' +\\
  \left[ \lambda z(1-\lambda z) +(x-\lambda^2)^2 - \lambda^2-\mu^2
  \right]g=0.
\end{multline}
So, fixing values $G_\sigma(z,p)$ and $G'_\sigma(z,p)$ at a given $z$, we 
determine $G_\sigma(z,p)$ uniquely. Hence, the spectrum of the Rabi
problem in Braak's settings is determined by \emph{two} transcendental
equation $G_\sigma(z_{\star},p)=0$ and $G'_\sigma(z_{\star},p)=0$, not
just one $G_\sigma(z_{\star},p)=0$, for a certain $z_{\star}$, as is
claimed in \cite{Braak:11::}.  This is illustrated in
Fig.~\ref{fig:2}. Here we just remark that choice $z_{\star}=0$ made
in \cite{Braak:11::} was lucky because, for $x<30$, all zeros of
$G_\sigma(0,p)$ belong to the spectrum.

However, as we show later there exists just one transcendental
function such that $x$ belongs to the spectrum if and only if it is a
zero of this single function. This is a desirable property in a
general method for finding spectra of systems described in Bargmann
representation. An example of the Rabi model with broken
$\mathbb{Z}_2$ symmetry, given in the end of paper~\cite{Braak:11::}
shows that a direct generalisation of the Braak approach is not
obvious.

In the Bargmann representation many models are described by a system
of linear differential equations with rational coefficients. The
system depends on a spectral parameter $E$. The problem is to
distinguish all values $E_n$ of $E$ such that for $E=E_n$, the
considered system possesses a non-zero entire solution contained in
the Cartesian product of an appropriate number of Bargmann spaces
$\mathscr{H}$.

Here, for simplicity, we consider a system described by just one
equation of the second order
\begin{equation}
  \label{eq:s}
  \varphi'' + p(z) \varphi' + q(z) \varphi=0, 
\end{equation}
where $p(z)$ and $q(z)$ are rational functions $z$, which additionally
may depend rationally on some parameters.  We assume that the system
possesses regular singular points $s_1, \ldots, s_m\in\mathbb{C}$.
Infinity may be a regular or an irregular singular point.  Let
$\rho_i$ and $\sigma_i$ be the exponents at singular point $s_i$. We
want to formulate necessary conditions for the existence of a nonzero
entire solution of equation~\eqref{eq:s}.  Hence, let us assume that
$\varphi(z)$ is such a solution.  Thus, at each point $s_i$ one of
exponents, let us say $\rho_i$, is a non-negative integer, and a local
holomorphic solution at this point has the form
\begin{equation}
  \label{eq:l}
  \varphi_i(z):= \sum_{n=0}^\infty a_n^{(i)}(z-s_i)^{\rho_i+n}.
\end{equation}
The radius of convergence of this series is not smaller than
$r_i:=\min_{j\neq i}|s_i-s_j|$.  Coefficients $a_n^{(i)}$ can be
computed recursively.

To proceed we have to make the following technical assumption. Namely,
we require that
\begin{equation}
  \label{eq:inter}
  U_i:= D(s_i,r_i)\cap
  D(s_{i+1},r_{i+1})\neq\emptyset,
\end{equation}
for $i=1,\ldots, m-1$. If none of $\sigma_i$ is a non-negative
integer, then, up to a multiplicative constant, series~\eqref{eq:l}
are just Taylor expansions of entire solution $\varphi(z)$ at points
$s_i$.  This is why local solutions $\varphi_{i}(z)$ and
$\varphi_{i+1}(z)$ have to be proportional for all $z\in U_i$ where both 
 are well defined. Hence, for a certain nonzero
$\alpha_i\in\mathbb{C}$, the function $F_i(z):=
\varphi_i(z)-\alpha_i\varphi(z)$ vanishes for all $z\in U_{i}$, so it
vanishes identically.  But $F_i(z)$, as a linear combination of
solutions of equation~\eqref{eq:s}, is its solution. So, $F_{i}(z)$
vanishes identically if and only if $F_{i}(z_\star)=0$, and
$F'_i(z_\star)=0$, \emph{for an arbitrary} $z_\star \in U_i$.
Eliminating $\alpha_i$ from equations $F_i(z_\star)=0$ and
$F'_i(z_\star)=0$, we obtain the following criterion. If, with the
given assumptions, equation~\eqref{eq:s} has a non-zero entire
solution, then
\begin{equation}
  \label{eq:w}
  W_i(z_i):=\det\begin{bmatrix}
    \varphi_i(z_i) &  \varphi_{i+1}(z_i) \\ 
    \varphi_i'(z_i) &  \varphi_{i+1}'(z_i) 
  \end{bmatrix}=0,
\end{equation}
for a certain $z_i \in U_i$, and $i=1,\ldots, m-1$. These $(m-1)$
conditions guarantee that $\varphi(z)$ is an entire function.  If infinity is a
regular singular point, they guarantee also that $\varphi(z)$ belongs
to the Bargmann space.

Formulation of this method with relaxed assumptions and its version
valid for systems of differential equations will be published in
\cite{kus:12::f}.

\begin{figure}[ht]
  \begin{center}
    \includegraphics[width=1\columnwidth,clip]{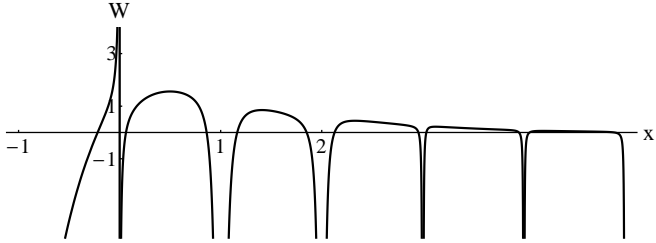}
  \end{center}
  \caption{\label{fig:3} Graph of Wronskian $W(p)$ for
    $p:=(x,\lambda,\mu,\epsilon)=(x,7/10, 4/10,0)$.}
\end{figure}

\begin{figure}[ht]
  \begin{center}
    \includegraphics[width=0.9\columnwidth,clip]{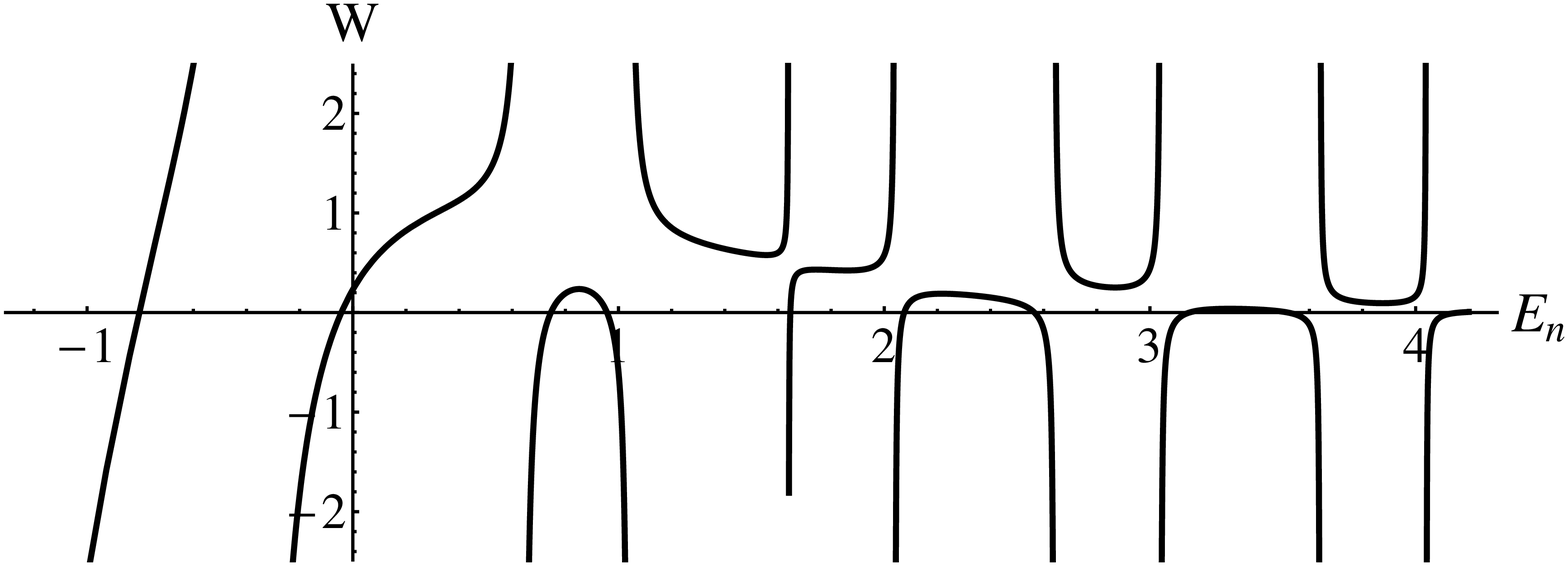}
  \end{center}
  \caption{\label{fig:4} Graph of Wronskian $W(p)$ for
    $p:=(E,\lambda,\mu,\epsilon)=(E,7/10, 4/10,2/10)$.}
\end{figure}

\begin{figure}[ht]
  \begin{center}
    \includegraphics[width=1\columnwidth,clip]{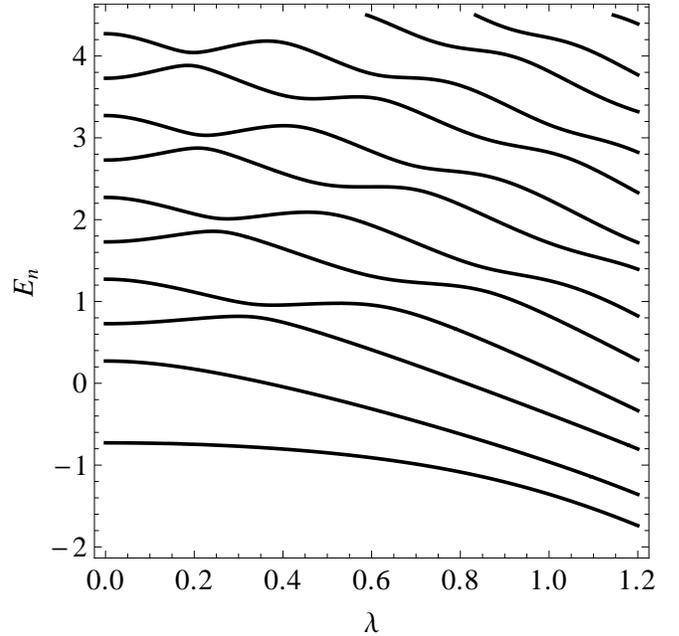}
  \end{center}
  \caption{\label{fig:5} Spectrum of generalised Rabi model for
    $\mu=0.7$, and $\epsilon=0.2$.}
\end{figure}

Let us consider the $\mathbb{Z}_{2}$ symmetry-broken Rabi
model~\eqref{eq:syste}. An elimination of $\psi_2(z)$ from this system
gives one second order equation for $\varphi(z):=\psi_1(z)$. This
equation has the form~\eqref{eq:s} with
\begin{equation}
  \begin{split}
    p(z)&=- \dfrac{\lambda + 2 \epsilon \lambda +
      z (-1 + 2 E + 2 \lambda^2)}{z^2 - \lambda^2}\\
    q(z)&=-\dfrac{\epsilon^2-E^2 + 2 z \epsilon \lambda + \lambda
      (\lambda + z (-1 + z \lambda)) + \mu^2}{z^2 - \lambda^2}.
  \end{split}
\end{equation}
Then, we introduce variables $y$ and $v(y)$ setting $z=\lambda(2y-1)$,
and
\begin{equation}
  \label{eq:vy}
  v(y):=\exp(-2\lambda^2y)\psi_1(\lambda(2y-1)). 
\end{equation}
By direct computation we obtain that $v(y)$ satisfies the following
equation
\begin{equation}
  v''+\left(\alpha+\dfrac{\beta+1}{y}+\dfrac{\gamma+1}{y-1}
  \right)v'+\left(\dfrac{ \widetilde\mu }{
      y}+\dfrac{\widetilde\nu}{y-1}\right)v=0,
  \label{eq:Heun}
\end{equation} 
where
\[
\begin{split}
  \widetilde\mu&=\dfrac{1}{2}(\alpha-\beta-\gamma+\alpha\beta-\beta\gamma)-\eta,
  \\
  \widetilde\nu&=\dfrac{1}{2}
  (\alpha+\beta+\gamma+\alpha\gamma+\beta\gamma)+\delta+\eta,
\end{split}
\]
and parameters $a_0=(\alpha,\beta,\gamma,\delta,\eta)$ were defined earlier, see~\eqref{eq:par_heun_def}. Equation~\eqref{eq:Heun} is
the Heun confluent equation. Its local holomorphic solution near $y=0$
is given by the Heun confluent function
$v_0(y)=\operatorname{HeunC}(a_0 ; y) $. Solution holomorphic near
$y=1$ is given by $ v_1(y)=\operatorname{HeunC}(a_1,1-y)$, where
$a_{1}:=(-\alpha,\gamma,\beta,-\delta,\delta+\eta)$.  Hence, local
solutions $\varphi_{\mp}(z)$ holomorphic near singular points
$z=\mp\lambda$ are given by
\begin{equation}
  \label{eq:ps1-}
  \varphi_{-}(z):=\exp[\lambda z+\lambda^2]\operatorname{HeunC}(a_0; ( 1+z/\lambda)/2)
\end{equation}
and
\begin{equation}
  \label{eq:ps1+}
  \varphi_{+}(z):=\exp[\lambda z+\lambda^2]\operatorname{HeunC}(a_1; ( 1-z/\lambda)/2).
\end{equation}
These local solutions are given by expansion of one entire function,
and hence their Wronskian
\begin{equation}
  W(p) := \varphi_+(z_\star)\varphi'_-(z_\star) - 
  \varphi'_+(z_\star)\varphi_-(z_\star),
\end{equation}
must vanish at arbitrary $z_\star$. The graph of the Wronskian $W(p)$
as a function of energy for the Rabi model and the generalised Rabi
model are shown in Fig.~\ref{fig:3}, and Fig.~\ref{fig:4},
respectively.  The resulting spectrum as seen in Fig.~\ref{fig:5}
reproduces that of \cite{Braak:11::}, with the benefit of not
including possible isolated zeroes of $G_+$.

This research has been partially supported by grant No.
DEC-2011/02/A/ST1/00208 of National Science Centre of Poland.
%
\def\polhk#1{\setbox0=\hbox{#1}{\ooalign{\hidewidth
  \lower1.5ex\hbox{`}\hidewidth\crcr\unhbox0}}} \def\cprime{$'$}
  \def\cydot{\leavevmode\raise.4ex\hbox{.}} \def\cprime{$'$}

\end{document}